\documentclass[11pt]{article}

\usepackage{amsmath,amssymb,amsthm}

\newtheorem{proposition}{Proposition}
\newtheorem{theorem}[proposition]{Theorem}
\newtheorem{lemma}[proposition]{Lemma}
\newtheorem{corollary}[proposition]{Corollary}
\theoremstyle{remark}
\newtheorem{remark}[proposition]{Remark}

\numberwithin{equation}{section}

\newcommand{\Hs}{\mathcal{H}}
\newcommand{\Ps}{\mathcal{P}}
\newcommand{\Sp}{\mathcal{S}}
\newcommand{\CP}{\mathbb{C}P}
\newcommand{\C}{\mathbb{C}}
\newcommand{\R}{\mathbb{R}}
\newcommand{\bra}[1]{\langle #1|}
\newcommand{\ket}[1]{|#1\rangle}
\newcommand{\braket}[2]{\langle #1|#2\rangle}
\newcommand{\ev}[1]{\langle #1\rangle}
\newcommand{\FS}{\mathrm{FS}}
\newcommand{\tr}{\operatorname{tr}}
\newcommand{\Span}{\operatorname{span}}

\title{Quantum Superposition Principle and Geometry}
\author{Alejandro Corichi\thanks{Electronic address: \texttt{corichi@matmor.unam.mx}}\\[4pt]
{\small Centro de Ciencias Matem\'aticas, Universidad Nacional Aut\'onoma de M\'exico,}\\
{\small A. Postal 61-3, Morelia, Michoac\'an 58089, M\'exico}}
\date{}

\begin{document}
\maketitle

\begin{abstract}
If one takes seriously the postulate that physical states are rays in Hilbert space, one is led to the geometric formulation of quantum mechanics, in which the space of states $\Ps$ is a K\"ahler manifold and the postulates of the theory are expressed through a symplectic form $\omega$ and a Riemannian metric $g$. The superposition principle does not fit naturally into this picture, since $\Ps$ is not a linear space. We argue that the principle should be split into two distinct principles with different physical and mathematical origins. The \emph{projective superposition principle} states that any two states determine a canonical two-sphere in $\Ps$, the projective line through them, whose points are their superpositions. We show that every geodesic of $\Ps$ through two states lies in this sphere, that it is the unique area-minimizing surface of its kind through them, and that a symplectic flow on $\Ps$ maps such spheres onto spheres if and only if it is generated by a quantum (quadratic) Hamiltonian, when $\dim\Hs\ge 3$. Projective superposition is thus equivalent to the linearity of quantum dynamics. The \emph{decomposition principle}, exemplified by the two-slit experiment, concerns the decomposition of a state by orthogonal projectors; here linearity lives in the space of observables, not of states. We give a fully geometric expression of the resulting interference pattern: the contribution of each slit is a product of transition probabilities set by geodesic distances, while the interference phase is the symplectic area, in units of $\hbar$, of a geodesic quadrilateral joining source, slits and detector. Finally, we show that the Robertson--Schr\"odinger uncertainty relation is a direct consequence of the compatibility between $g$ and $\omega$, and clarify the geometric status of the canonical commutation relations.
\end{abstract}

\begingroup\small
\noindent\textbf{Note on this version.} This version (v4) is a thoroughly revised and extended version of v3 (2006) of this article~\cite{v1}; a version of the paper appeared in Gen.\ Rel.\ Grav.\ \textbf{38}, 677--687 (2006). Normalization conventions have been made consistent, the uncertainty relation of the former Appendix~B (stated incorrectly there) has been corrected, the proof of the totally geodesic property has been replaced by a shorter and more complete one, the uniqueness statement for the sphere spanned by two states has been made precise, and the statement that eigenstates are local extrema of observables has been corrected. Sections~\ref{sec:dynamics} and~\ref{sec:interference} and Proposition~\ref{prop:luders} are new.
\par\endgroup

\section{Introduction}\label{sec:intro}

It has been known for some time that quantum mechanics, with all its postulates, can be formulated in geometric language~\cite{Kibble,Heslot,CMP,AS,BH,BZ}. The starting point is the observation that physical states are not vectors in the Hilbert space $\Hs$ but rays, so the true space of states is the projective space $\Ps$. This space carries a natural K\"ahler structure: a symplectic form $\omega$, with respect to which Schr\"odinger evolution is a Hamiltonian flow exactly as in classical mechanics, and a Riemannian metric $g$, which encodes the specifically quantum features of the theory, namely transition probabilities and uncertainties~\cite{AS,BH,AA90}.

In most treatments of the geometric formulation the superposition principle is not discussed. The obvious reason is that $\Ps$ is not a linear space: the sum of two states is not defined. Given the prominent place the principle holds in standard presentations of the theory~\cite{Dirac,Feynman}, this may seem to signal an incompatibility between the Hilbert-space and the geometric formulations. The purpose of this paper is to show that there is no such incompatibility, and that the geometric formulation in fact teaches us something about what the superposition principle says.

Our central claim is that the superposition principle, as normally stated, conflates two distinct principles:
\begin{itemize}
\item[(i)] a \emph{projective superposition principle}, by which any two states determine a canonical family of states, their superpositions, forming a totally geodesic two-sphere in $\Ps$; and
\item[(ii)] a \emph{decomposition principle}, by which a state is decomposed by a family of orthogonal projectors, as happens when a beam traverses a screen with several slits.
\end{itemize}
In (i) one is free to prepare any point of the sphere, as in a Stern--Gerlach experiment. In (ii) the incoming state is fixed and the freedom lies in the projectors; the linear structure that produces interference is that of the space of observables, not that of the state space.

The two principles rest on different geometric structures of $\Ps$, and each has a sharp consequence. For (i), we show that requiring a symplectic flow to preserve projective superpositions forces it to be generated by a Hermitian operator: the linearity of the Schr\"odinger equation is equivalent to the dynamical stability of projective superpositions (Theorem~\ref{thm:dynamics}). For (ii), we show that the two-slit interference pattern is entirely geometric. The contribution of each slit is a product of transition probabilities, determined by $g$, and the interference phase is a symplectic area in units of $\hbar$, determined by $\omega$ (Theorem~\ref{thm:interference}).

The paper is organized as follows. Section~\ref{sec:geometry} reviews the geometric formulation with consistent normalizations. Section~\ref{sec:superposition} develops the projective superposition principle and the geometry of the spheres spanned by pairs of states. Section~\ref{sec:dynamics} relates it to the linearity of the dynamics. Section~\ref{sec:decomposition} discusses the decomposition principle and Section~\ref{sec:interference} its geometric interference formula. Section~\ref{sec:uncertainty} discusses uncertainty relations, and Section~\ref{sec:discussion} contains a discussion. Proofs of the more technical statements are collected in the appendices.

\section{The geometric formulation of quantum mechanics}\label{sec:geometry}

We work with a finite-dimensional Hilbert space $\Hs\simeq\C^n$; most statements extend to the infinite-dimensional case with the usual care about domains of unbounded operators~\cite{CMP,AS}. Regard $\Hs$ as a real vector space with complex structure $J$ (multiplication by $i$) and decompose the Hermitian inner product, antilinear in the first entry, into real and imaginary parts:
\begin{equation}\label{eq:inner}
\braket{\psi}{\phi} = \frac{1}{2\hbar}\Big[G(\psi,\phi) + i\,\Omega(\psi,\phi)\Big].
\end{equation}
Here $G$ is a Euclidean inner product and $\Omega$ a symplectic form on $\Hs$, related by $\Omega(X,Y)=G(JX,Y)$. The factor $1/2\hbar$ is chosen so that Schr\"odinger evolution is generated by the expectation value of the Hamiltonian; we follow the conventions of~\cite{AS}.

Let $\Sp$ be the unit sphere in $\Hs$. The space of states is the quotient $\Ps=\Sp/U(1)=\CP^{n-1}$, and we write $\Psi=[\psi]$ for the state of a unit vector $\psi$. At $\psi\in\Sp$ a tangent vector $X$ is \emph{horizontal} if $\braket{\psi}{X}=0$; horizontal vectors are identified with tangent vectors to $\Ps$ at $[\psi]$, and $X_h:=X-\braket{\psi}{X}\psi$ denotes the horizontal part of $X$. Since $G$ and $\Omega$ are invariant under phase rotations, their restrictions to horizontal vectors define a metric $g$ and a two-form $\omega$ on $\Ps$:
\begin{equation}
g(X,Y)=G(X_h,Y_h),\qquad \omega(X,Y)=\Omega(X_h,Y_h).
\end{equation}
$(\Ps,g,\omega,J)$ is a K\"ahler manifold, with $\omega(X,Y)=g(JX,Y)$. It is often convenient to use the Fubini--Study normalizations
\begin{equation}\label{eq:FS}
g_\FS := \frac{g}{2\hbar} = \mathrm{Re}\braket{X_h}{Y_h},\qquad \omega_\FS:=\frac{\omega}{2\hbar}=\mathrm{Im}\braket{X_h}{Y_h},
\end{equation}
which do not involve $\hbar$. With $g_\FS$ the space $\CP^1$ is a round sphere of radius $1/2$ and area $\pi$.

\paragraph{Observables and dynamics.} To each Hermitian operator $F$ we associate the real function $f(\Psi)=\ev{F}=\bra{\psi}F\ket{\psi}$ on $\Ps$. Its Hamiltonian vector field, defined by $\iota_{X_f}\omega=df$, is the projection to $\Ps$ of the generator of the unitary group $e^{-iFt/\hbar}$:
\begin{equation}\label{eq:Xf}
X_f\big|_{[\psi]} = \Big(-\frac{i}{\hbar}F\psi\Big)_h = -\frac{i}{\hbar}\big(F-\ev{F}\big)\psi.
\end{equation}
The vector fields $X_f$ are Killing fields of $g$. Conversely, every function whose Hamiltonian vector field is Killing is of the form $\ev{F}$ for some Hermitian $F$~\cite{AS}. The map $F\mapsto X_f$ is real-linear, with kernel the multiples of the identity. Schr\"odinger evolution is the Hamiltonian flow of $h=\ev{H}$. The Poisson bracket $\{f,m\}:=\omega(X_f,X_m)$ of two such observables satisfies
\begin{equation}\label{eq:PB}
\{f,m\}=\Big\langle\frac{1}{i\hbar}[F,M]\Big\rangle,\qquad \dot f=\{f,h\},\qquad [X_f,X_m]=-X_{\{f,m\}},
\end{equation}
the second identity being Ehrenfest's theorem written as Hamilton's equations. So far everything is as in classical mechanics, except that the quantum observables form a very small subset of the functions on $\Ps$.

\paragraph{The metric: uncertainties and probabilities.} The metric $g$ has no classical counterpart. Define the symmetric product of two observables
\begin{equation}\label{eq:cov}
(f,m):=\frac{\hbar}{2}\,g(X_f,X_m)=\tfrac12\ev{FM+MF}-\ev{F}\ev{M},
\end{equation}
so that $(f,f)=(\Delta F)^2$: the length of $X_f$ is the uncertainty of $F$. The zeros of $X_f$ are the critical points of $f$, and by~\eqref{eq:Xf} these are precisely the eigenstates of $F$, the critical values being the eigenvalues. Only the eigenstates of the largest and smallest eigenvalues are extrema of $f$; the others are saddle points.

Transition probabilities are given by geodesic distances. Writing $d$ for the distance of $g$ and $\theta=d/\sqrt{2\hbar}$ for the distance of $g_\FS$ (the \emph{Fubini--Study angle}), one has, for unit vectors,
\begin{equation}\label{eq:prob}
|\braket{\psi}{\phi}|^2=\cos^2\theta(\Psi,\Phi)=\cos^2\!\Big(\frac{d(\Psi,\Phi)}{\sqrt{2\hbar}}\Big),\qquad 0\le\theta\le\frac{\pi}{2}.
\end{equation}
A proof is given in Proposition~\ref{prop:sphere} below. The relation~\eqref{eq:prob} also covers degenerate measurements:

\begin{proposition}\label{prop:luders}
Let $E$ be the spectral projector of an eigenvalue $\lambda$ of $F$, and let $\Ps(E)\subset\Ps$ be the set of states in its range, a totally geodesic submanifold isometric to $\CP^{k-1}$, $k=\tr E$. If the system is in the state $\Psi$, the probability of obtaining $\lambda$ is
\begin{equation}
\mathrm{Prob}(\lambda\,|\,\Psi)=\bra{\psi}E\ket{\psi}=\cos^2\theta\big(\Psi,\Ps(E)\big),
\end{equation}
where $\theta(\Psi,\Ps(E))$ is the distance from $\Psi$ to the submanifold $\Ps(E)$. If $E\psi\neq 0$, the point of $\Ps(E)$ nearest to $\Psi$ is unique and equals $[E\psi]$, the L\"uders post-measurement state~\cite{Luders}.
\end{proposition}

\begin{proof}
For unit $\phi$ in the range of $E$, $|\braket{\phi}{\psi}|=|\braket{\phi}{E\psi}|\le\|E\psi\|$ by the Cauchy--Schwarz inequality, with equality iff $\phi\propto E\psi$. Hence $\min_{\Phi\in\Ps(E)}\theta(\Psi,\Phi)=\arccos\|E\psi\|$, attained only at $[E\psi]$. The totally geodesic property is proved in Appendix~\ref{app:geodesic}.
\end{proof}

A state is therefore more likely to ``collapse'' onto a nearby eigenspace than onto a distant one, and the collapse is the nearest-point projection. This ends our review of the geometric formulation.

\paragraph{The problem.} The space $\Ps$ is not linear: for $\psi'=e^{i\alpha}\psi$ one has $[\psi+\phi]\neq[\psi'+\phi]$ in general, so the sum of vectors does not descend to $\Ps$. In the following sections we re-examine the superposition principle in view of this fact.

\section{The projective superposition principle}\label{sec:superposition}

In $\Hs$ the superposition of two vectors, $\alpha\psi+\beta\phi$, combines the two linear operations, sum and multiplication by scalars. For fixed $\psi,\phi$ the set of all superpositions is the two-dimensional subspace $\Span\{\psi,\phi\}\simeq\C^2$. Although states cannot be added, the following construction generalizes linear combination to $\Ps$.

Let $\Psi\neq\Phi$ be two states and choose unit representatives $\psi$, $\phi$ with $\braket{\psi}{\phi}=0$. If the given states are not orthogonal, replace $\phi$ by the normalized component of $\phi$ orthogonal to $\psi$; the original $\Phi$ then corresponds to a finite value of $z$ below, and the set of superpositions is unchanged. For $z\in\C\cup\{\infty\}$ define the \emph{projective superposition}
\begin{equation}\label{eq:projsup}
\Psi\overset{z}{\oplus}\Phi := [\,\psi+z\,\phi\,],\qquad \Psi\overset{\infty}{\oplus}\Phi:=\Phi .
\end{equation}
The set of all such states is
\begin{equation}
L_{\Psi\Phi}:=\{\Psi\overset{z}{\oplus}\Phi\;:\;z\in\C\cup\{\infty\}\}=\Ps\big(\Span\{\psi,\phi\}\big)\simeq\CP^1,
\end{equation}
the \emph{projective line} through $\Psi$ and $\Phi$. It depends only on the two states, not on the representatives. Topologically it is a sphere on which $z$ is a stereographic (Riemann) coordinate, with $\Psi$ at the origin and $\Phi$ at infinity. Writing $z=\tan(\beta/2)\,e^{i\alpha}$, the angles $(\beta,\alpha)$ are the polar and azimuthal angles of the Bloch sphere with $\Psi$ at the north pole, and
\begin{equation}
|\braket{\psi}{\Psi\overset{z}{\oplus}\Phi}|^2=\frac{1}{1+|z|^2}=\cos^2\frac{\beta}{2},\qquad \theta\big(\Psi,\Psi\overset{z}{\oplus}\Phi\big)=\frac{\beta}{2}.
\end{equation}
Thus there is ``a sphere's worth'' of superpositions of two given states.

\paragraph{The relative phase.} The modulus $|z|$ is determined by the transition probabilities. The phase $\arg z$ is not determined by $\Psi$ and $\Phi$ alone: replacing $\phi$ by $e^{i\lambda}\phi$ sends $z\mapsto e^{-i\lambda}z$, a rotation of the Bloch sphere about the axis through $\Psi$ and $\Phi$. Replacing $\psi$ by $e^{i\lambda}\psi$ produces the rotation in the opposite sense. This is \emph{not} a change of complex structure, which is fixed on $\Ps$; it is the freedom of choosing the meridian $\arg z=0$. The relative phase of a superposition is therefore meaningful only with respect to a third reference state $\Xi$ on the line, declared to have $z=1$. Given $\Xi$, the coordinate $z$ of a state $X=[\chi]$, $\chi=\psi+z\phi$, is fixed by representative-independent data: the transition probabilities among $\Psi$, $\Xi$, $X$ and the phase of the Bargmann invariant
\begin{equation}
\arg\Big(\braket{\psi}{\xi}\braket{\xi}{\chi}\braket{\chi}{\psi}\Big)=\arg(1+z),\qquad \xi=\psi+\phi .
\end{equation}
This is the geometric counterpart of the fiducial vectors used in~\cite{MMSZ} to describe superpositions in terms of density operators. Section~\ref{sec:interference} shows that such invariants are symplectic areas.

\paragraph{The geometry of the lines.} The following proposition collects the geometric properties of the lines; its proof is in Appendix~\ref{app:geodesic}.

\begin{proposition}\label{prop:sphere}
Let $\Psi\neq\Phi$ be states in $\Ps$.
\begin{enumerate}
\item[(a)] $L_{\Psi\Phi}$ is the unique projective line (complex projective curve of degree one) containing $\Psi$ and $\Phi$.
\item[(b)] Every geodesic of $(\Ps,g)$ passing through $\Psi$ and $\Phi$ lies entirely in $L_{\Psi\Phi}$. In particular $L_{\Psi\Phi}$ is totally geodesic, and the distance between $\Psi$ and $\Phi$ in $\Ps$ equals their distance on the round sphere $L_{\Psi\Phi}$, which gives~\eqref{eq:prob}.
\item[(c)] With respect to $g_\FS$, $L_{\Psi\Phi}$ has area $\pi$. Among all closed oriented surfaces in $\Ps$ that represent the generator of $H_2(\Ps;\mathbb{Z})$ and contain $\Psi$ and $\Phi$, it is the unique one of minimal area.
\end{enumerate}
More generally, if $W$ is the span of $k+1$ linearly independent vectors, the projective subspace $\Ps(W)\simeq\CP^k$ is totally geodesic.
\end{proposition}

Part (c) makes precise the statement in~\cite{v1} that the sphere is singled out by its normalization. Part (b) is perhaps the most surprising. Transition probabilities between arbitrary states of an arbitrarily large system are computed on a single qubit sphere, the one spanned by the two states, because the geodesic joining them never leaves it. In this sense the qubit, a sphere of radius $1/2$, is the basic building block of the quantum state space.

\paragraph{Physical realization.} The projective superposition principle is realized in the preparation of a spin-$\tfrac12$ system. With a Stern--Gerlach apparatus oriented along any direction in space one prepares any point of the sphere $L_{\Psi\Phi}$, $\Psi=[\ket{+}]$, $\Phi=[\ket{-}]$, and this orientation is the only choice available in the preparation. The physical freedom of preparation is thus in one-to-one correspondence with the points of the projective line. A related nonlinear formulation of the superposition principle was developed independently in~\cite{CGM1,CGM2}. More recently, within a programme that formulates quantum theory in terms of complex projective geometry~\cite{Sontz1,Sontz2}, Sontz~\cite{Sontz3} has proposed to \emph{define} the superposition of two distinct states as the projective line through them, i.e.\ as $L_{\Psi\Phi}$ regarded as a single object. He extends the definition to arbitrary families of states by taking the smallest projective subspace that contains them, which is the subspace $\CP^k$ of Proposition~\ref{prop:sphere}. This is the projective superposition principle discussed here (and already in~\cite{v1}). The emphasis of the present paper is on the individual points of the line, on their parametrization by $z$, and on the phase reference needed to single out one of them.

\section{Projective superposition and the linearity of dynamics}\label{sec:dynamics}

In the Hilbert-space formulation, linearity has two roles: the state space is linear, and the dynamics is linear. In the geometric formulation the first is absent. We now show that the second is equivalent to the stability of projective superpositions under evolution. Note that any Hamiltonian flow on $(\Ps,\omega)$ is a candidate dynamics, and that flows of functions other than $\ev{H}$ (e.g.\ Weinberg's nonlinear quantum mechanics~\cite{Weinberg}) are symplectic but are not isometries of $g$.

\begin{theorem}\label{thm:dynamics}
Let $\dim\Hs=n\ge3$ and let $T:\Ps\to\Ps$ be a continuous bijection that maps projective lines onto projective lines and preserves $\omega$. Then $T$ is induced by a unitary operator: $T([\psi])=[U\psi]$.
\end{theorem}

\begin{proof}
By the fundamental theorem of projective geometry~\cite{Artin}, a bijection of $\Ps(\C^n)$, $n\ge3$, that maps lines onto lines is induced by a semilinear bijection $A$ of $\C^n$, i.e.\ $A(\lambda v)=\sigma(\lambda)A(v)$ for some field automorphism $\sigma$ of $\C$. Since $T$ is continuous, so is $\sigma$, hence $\sigma$ is the identity or complex conjugation. In the second case $T$ is antiholomorphic, $dT\circ J=-J\circ dT$, and if $T^*\omega=\omega$ then, using $g(X,Y)=\omega(X,JY)$,
\[
g(dT X,dT Y)=\omega(dTX,J\,dTY)=-\omega(dTX,dT\,JY)=-g(X,Y),
\]
which is impossible for a Riemannian metric. Hence $A$ is complex linear, $T$ is holomorphic, and by the same identity $T$ is an isometry of $g$. An isometry preserves the transition probabilities~\eqref{eq:prob}, so by Wigner's theorem~\cite{Wigner,Bargmann} it is induced by a unitary or an antiunitary operator; being holomorphic, it is induced by a unitary one.
\end{proof}

\begin{corollary}\label{cor:linear}
Let $\dim\Hs\ge3$ and let $h$ be a smooth function on $\Ps$ whose Hamiltonian flow maps projective lines onto projective lines. Then $h=\ev{H}+c$ for a Hermitian operator $H$ and a constant $c$. That is, the flow is the Schr\"odinger evolution generated by $H$.
\end{corollary}

\begin{proof}
By Theorem~\ref{thm:dynamics} the flow $T_t$ is a one-parameter group in $PU(n)=U(n)/U(1)$. Since the Lie algebra map $\mathfrak{u}(n)\to\mathfrak{pu}(n)$ is onto, it lifts to a one-parameter unitary group $e^{-iHt/\hbar}$. Then $X_h=X_{\ev{H}}$, so $d(h-\ev{H})=0$, and $\Ps$ is connected.
\end{proof}

The \emph{dynamical content} of the projective superposition principle is therefore this: if $\Psi$ evolves into $\Psi'$ and $\Phi$ into $\Phi'$, every superposition of $\Psi$ and $\Phi$ evolves into a superposition of $\Psi'$ and $\Phi'$. Among all Hamiltonian dynamics on $\Ps$, this requirement singles out the linear Schr\"odinger equation. Nonlinear modifications bend the lines. For instance, the flow of $h=\ev{H}+\varepsilon\ev{E}^2$, with $E$ a projector, does not: for generic $H$ and $\varepsilon\neq0$, the images of three points of a line no longer lie on a common line. Two remarks are in order. First, for a qubit ($n=2$) there is only one line and the hypothesis is empty, exactly as Gleason's theorem~\cite{Gleason} and Uhlhorn's version of Wigner's theorem~\cite{Uhlhorn} require $n\ge3$. For a qubit, linearity must be imposed through the Killing condition on $X_h$. Second, other arguments for linearity rely on composite systems and the absence of superluminal signalling~\cite{Gisin,Polchinski}; Corollary~\ref{cor:linear} instead refers to a single system and to the structure of its state space only.

\paragraph{An Erlangen reading.} Theorem~\ref{thm:dynamics} can also be read in the spirit of Klein's Erlangen programme. The projective geometry of $\Ps$, defined by its points and lines, has as symmetry group the continuous collineations: for $n\ge3$, the group $PGL(n,\C)$ together with its composition with complex conjugation. The geometry relevant to quantum theory is defined instead by the transformations that preserve transition probabilities, i.e.\ by the isometries of $g$: the unitary and antiunitary maps (Wigner's theorem). Sontz~\cite{Sontz3} argues that the appropriate group is $PU(n)$, the antiunitary maps being excluded because time reversal is not a universal symmetry. Theorem~\ref{thm:dynamics} characterizes this group without reference to probabilities: the continuous collineations that preserve $\omega$ are exactly the elements of $PU(n)$. An antiunitary map preserves the lines and the metric but reverses the sign of $\omega$, i.e.\ it is anti-symplectic. This is the geometric reason why time reversal, when it is a symmetry, is represented antiunitarily: it reverses the orientation of time and with it the symplectic form, exactly as $t\mapsto-t$, $p\mapsto-p$ does in classical mechanics.

\section{The decomposition principle}\label{sec:decomposition}

One of the main differences between the classical and the quantum descriptions concerns how probabilities are computed. Classically, the probabilities of mutually exclusive alternatives leading to the same outcome are added. In quantum mechanics one adds complex amplitudes and computes the modulus squared at the end, which brings in interference. The paradigmatic example is the two-slit experiment, which we now analyse within the geometric description. We want to distinguish it from the superposition of states, be it linear or projective, and for this reason we call the relevant principle the \emph{decomposition principle}.

A beam prepared in the state $\Psi=[\psi]$ traverses a wall with two slits and is detected on a screen. Following Feynman~\cite{Feynman}, the wall is represented by orthogonal projectors $E_1,E_2$ ($E_iE_j=\delta_{ij}E_j$) onto the states that pass through each slit, and the wall as a whole by $E_{\rm w}=E_1+E_2$, which is again a projector. We describe detection by a unit vector $\chi$; for a point $x$ on the screen, $\chi$ is a narrow wave packet centred at $x$, and position eigenstates are recovered in the limit. Given that the particle has passed the wall, the probability of detection is the ray-invariant function on $\Ps$
\begin{equation}\label{eq:twoslit}
p(\chi\,|\,\Psi)=\frac{|\bra{\chi}E_{\rm w}\ket{\psi}|^2}{\bra{\psi}E_{\rm w}\ket{\psi}},
\end{equation}
and since $\bra{\chi}E_{\rm w}\ket{\psi}=\bra{\chi}E_1\ket{\psi}+\bra{\chi}E_2\ket{\psi}=:\varphi_1+\varphi_2$, the familiar interference term $2\,\mathrm{Re}(\varphi_1\overline{\varphi_2})$ appears.

The important point is that $\varphi_1\overline{\varphi_2}$ is invariant under $\psi\mapsto e^{i\alpha}\psi$, and after the normalization in~\eqref{eq:twoslit} the whole pattern is invariant under $\psi\mapsto c\,\psi$, $c\in\C^*$, so it is a function on $\Ps$. Moreover, the projectors $E_i$ are themselves defined by states: $E_i=\sum_k\ket{e_k}\bra{e_k}/\braket{e_k}{e_k}$ is insensitive to the phases of its basis vectors. The physical situation is thus very different from that of Section~\ref{sec:superposition}. In the two-slit experiment one cannot vary independently two states that are then combined. The incoming state is fixed, and the freedom of the experimenter lies in the wall: no wall, one slit, two slits, and so on.

\paragraph{Where the linearity lives.} The linear structure responsible for interference is that of the projectors, not that of the states. In the standard presentation the states appear to be superposed, but what one actually does is define the operator $E_{\rm w}$ as the sum of the projectors representing the alternatives, and then decompose the state with it. Geometrically, a Hermitian operator $F$ has two faces. First, its action on vectors induces a map $\Ps\setminus\Ps(\ker F)\to\Ps$, which in general preserves neither $\omega$ nor $g$. Second, it defines the Killing field $X_f$ of~\eqref{eq:Xf}, the infinitesimal generator of the symmetries $e^{-iFt/\hbar}$. Killing fields of $(g,\omega)$ form a real vector space, and $F\mapsto X_f$ is linear, so $X_{e_1+e_2}=X_{e_1}+X_{e_2}$ with $e_i=\ev{E_i}$. Since $[E_1,E_2]=0$, the functions $e_i$ Poisson commute and the fields $X_{e_i}$ commute by~\eqref{eq:PB}. Projectors themselves do not form a vector space: if $E_1,E_2$ are nonzero orthogonal projectors, $\alpha E_1+\beta E_2$ is a projector only for $\alpha,\beta\in\{0,1\}$. Orthogonal projectors can only be \emph{added}, and this is how the decomposition principle enters. There is thus no inconsistency between the absence of a linear structure on $\Ps$ and the additivity of amplitudes in~\eqref{eq:twoslit}. The distinction between superposing states and expanding a given state in a basis has also been stressed in~\cite{Sontz3}, where the expansion is associated with the replacement of the state by a statistical mixture after a measurement. The decomposition principle concerns the opposite situation: the decomposition by the projectors of the wall is coherent, no measurement of the slit is performed, and the interference term survives.

\paragraph{The two principles meet.} Let $a_i:=E_i\psi/\|E_i\psi\|$, assuming $E_i\psi\neq0$. By Proposition~\ref{prop:luders}, $A_i=[a_i]$ is the point of the ``slit submanifold'' $\Ps(E_i)$ nearest to $\Psi$. The state after the wall is
\begin{equation}\label{eq:postwall}
[E_{\rm w}\psi]=\big[\,\|E_1\psi\|\,a_1+\|E_2\psi\|\,a_2\,\big]=A_1\overset{z}{\oplus}A_2,\qquad z=\frac{\|E_2\psi\|}{\|E_1\psi\|}>0,
\end{equation}
a projective superposition of $A_1$ and $A_2$. The relative-phase ambiguity of Section~\ref{sec:superposition} is absent here: the representatives $a_i$ are fixed by the condition $\braket{a_i}{\psi}>0$, i.e.\ the incoming state itself provides the phase reference. This explains, in geometric terms, why the interference pattern does not depend on the phase of $\psi$. The two principles are not independent, but they assign the physical freedom differently. In projective superposition one chooses a point of a line. In decomposition, the line and the point on it are produced by the projectors from a single given state.

\section{The geometry of interference}\label{sec:interference}

The original version of this paper~\cite{v1} left open how to visualize the two-slit interference in terms of the geometric objects available on $\Ps$. The answer is that the moduli in~\eqref{eq:twoslit} are governed by $g$ and the phase by $\omega$.

\begin{theorem}\label{thm:interference}
In the setting of Section~\ref{sec:decomposition}, assume that $\braket{\chi}{a_i}\neq0$ for $i=1,2$. Then
\begin{equation}\label{eq:fringe}
p(\chi\,|\,\Psi)\,\bra{\psi}E_{\rm w}\ket{\psi}=p_1+p_2+2\sqrt{p_1p_2}\,\cos\Phi,
\end{equation}
where
\begin{equation}\label{eq:pi}
p_i=|\bra{\chi}E_i\ket{\psi}|^2=\cos^2\theta(\Psi,A_i)\;\cos^2\theta(A_i,X)
\end{equation}
is the product of the transition probabilities from the source $\Psi$ to the nearest point $A_i$ of the slit submanifold, and from there to the detector state $X=[\chi]$. The interference phase is
\begin{equation}\label{eq:area}
\Phi=\frac{1}{\hbar}\int_\Sigma\omega = 2\int_\Sigma\omega_\FS \pmod{2\pi},
\end{equation}
where $\Sigma\subset\Ps$ is any surface whose oriented boundary is the geodesic quadrilateral $X\to A_1\to\Psi\to A_2\to X$ (see Fig.~\ref{fig:quad}).
\end{theorem}

\begin{figure}[t]
\centering
\setlength{\unitlength}{1cm}
\begin{picture}(9,5.4)(0,0)
  \thinlines
  \qbezier[300](1.1,3.4)(2.3,2.5)(2.1,1.1)
  \qbezier[300](7.9,3.4)(6.7,2.5)(6.9,1.1)
  \put(2.1,0.85){\makebox(0,0)[t]{$\Ps(E_1)$}}
  \put(6.9,0.85){\makebox(0,0)[t]{$\Ps(E_2)$}}
  \thicklines
  \qbezier(4.5,4.5)(3.2,4.0)(1.95,2.375)
  \qbezier(1.95,2.375)(3.0,1.2)(4.5,0.5)
  \qbezier(4.5,0.5)(6.0,1.2)(7.05,2.375)
  \qbezier(7.05,2.375)(5.8,4.0)(4.5,4.5)
  \put(4.5,4.5){\circle*{0.13}}
  \put(1.95,2.375){\circle*{0.13}}
  \put(4.5,0.5){\circle*{0.13}}
  \put(7.05,2.375){\circle*{0.13}}
  \put(4.5,4.7){\makebox(0,0)[b]{$X=[\chi]$ (detector)}}
  \put(1.75,2.375){\makebox(0,0)[r]{$A_1$}}
  \put(7.25,2.375){\makebox(0,0)[l]{$A_2$}}
  \put(4.5,0.3){\makebox(0,0)[t]{$\Psi=[\psi]$ (source)}}
  \put(4.5,2.8){\makebox(0,0){$\Sigma$}}
  \put(4.5,2.1){\makebox(0,0){$\Phi=\frac{1}{\hbar}\int_\Sigma\omega$}}
\end{picture}
\caption{Geometric picture of two-slit interference on the state space $\Ps$. The source state $\Psi$ is projected onto the nearest points $A_i$ of the slit submanifolds $\Ps(E_i)$ (Proposition~\ref{prop:luders}). The contribution of slit $i$ is the product of transition probabilities along $\Psi\to A_i\to X$, set by geodesic distances. The interference phase is the symplectic area enclosed by the geodesic quadrilateral, traversed as $X\to A_1\to\Psi\to A_2\to X$, in units of $\hbar$.}
\label{fig:quad}
\end{figure}

\begin{proof}
Since $\braket{a_i}{\psi}=\|E_i\psi\|>0$, one has $\varphi_i=\braket{\chi}{a_i}\braket{a_i}{\psi}$, which gives~\eqref{eq:pi}, and
\[
\varphi_1\overline{\varphi_2}=\braket{\chi}{a_1}\braket{a_1}{\psi}\braket{\psi}{a_2}\braket{a_2}{\chi}=\tr\big(\Pi_X\Pi_{A_1}\Pi_\Psi\Pi_{A_2}\big),
\]
where $\Pi$ denotes the rank-one projector onto a state. The right-hand side is a four-vertex Bargmann invariant~\cite{Bargmann}, independent of representatives, with modulus $\sqrt{p_1p_2}$. Its phase $\Phi$ is given by~\eqref{eq:area}: the phase of a Bargmann invariant equals $2\int\omega_\FS$ over any surface bounded by the corresponding geodesic polygon, provided consecutive vertices are not orthogonal~\cite{MS,SB,BZ}. For completeness we give the argument in Appendix~\ref{app:phase}. Note that $\braket{a_1}{\psi}$ and $\braket{\psi}{a_2}$ are nonzero by construction, and the remaining two factors are nonzero by hypothesis. Expanding $|\varphi_1+\varphi_2|^2$ gives~\eqref{eq:fringe}.
\end{proof}

Several remarks are in order.

\begin{remark}
The phase~\eqref{eq:area} is well defined. Two surfaces with the same boundary differ by a closed surface, and $\frac{1}{2\pi\hbar}\omega$ is an integral class, since $\int_{\CP^1}\omega=2\hbar\pi$ by Proposition~\ref{prop:sphere}(c). Hence $\Phi$ changes by multiples of $2\pi$. This is the same integrality that underlies geometric quantization and the Bohr--Sommerfeld conditions.
\end{remark}

\begin{remark}
Equation~\eqref{eq:fringe} separates the classical and the quantum aspects of the experiment in geometric terms. The terms $p_i$ are what one would obtain by treating ``passage through slit $i$'' as a classical intermediate event, and they depend only on the metric $g$. The interference term depends on the symplectic structure through an area, in close analogy with the Aharonov--Bohm phase as a magnetic flux. If the symplectic area vanishes, as happens when all four states lie on a common real slice, then $\cos\Phi=\pm1$ and the interference is purely constructive or destructive.
\end{remark}

\begin{remark}
Bargmann invariants are invariant under $PU(n)$ but not under the projective group $PGL(n,\C)$. In the terminology of~\cite{Sontz3}, they are quantum properties that are not properties of the standard projective geometry. Theorem~\ref{thm:interference} and Appendix~\ref{app:phase} explain why. The modulus of a Bargmann invariant is a product of cosines of geodesic distances, and its phase is a symplectic area. $PGL(n,\C)$ preserves neither $g$ nor $\omega$, whereas $PU(n)$ preserves both.
\end{remark}

\begin{remark}
With $N$ slits, $E_{\rm w}=\sum_iE_i$ and the same argument gives \[p\,\ev{E_{\rm w}}=\sum_ip_i+2\sum_{i<j}\sqrt{p_ip_j}\,\cos\Phi_{ij},\] with $\Phi_{ij}$ the area of the quadrilateral $X\to A_i\to\Psi\to A_j\to X$. Interference is therefore pairwise: in Sorkin's terminology~\cite{Sorkin} quantum theory has no third-order interference, $I_3=0$. This is a direct consequence of the decomposition principle together with the quadratic Born rule, and it has been tested experimentally~\cite{Sinha}. The absence of higher-order interference plays a central role in recent reconstructions of quantum theory from single-system postulates~\cite{BMU}.
\end{remark}

\section{Uncertainty relations and the K\"ahler structure}\label{sec:uncertainty}

The geometric formulation gives a transparent derivation of the uncertainty relations, which replaces the discussion in Appendix~B of~\cite{v1}. The key input is the compatibility between $g$ and $\omega$.

\begin{lemma}\label{lem:kahler}
On a K\"ahler manifold, for any two tangent vectors at a point,
\begin{equation}
g(X,X)\,g(Y,Y)\ \ge\ g(X,Y)^2+\omega(X,Y)^2 .
\end{equation}
\end{lemma}

\begin{proof}
If $X\neq0$, the vectors $X$ and $JX$ are orthogonal and of equal length, since $g(X,JX)=\omega(X,X)=0$ and $J$ is an isometry. The orthogonal projection of $Y$ onto their span has squared length $\big[g(X,Y)^2+g(JX,Y)^2\big]/g(X,X)$, which cannot exceed $g(Y,Y)$. Finally, $g(JX,Y)=\omega(X,Y)$.
\end{proof}

Applied to $X_f$ and $X_m$, and using~\eqref{eq:PB} and~\eqref{eq:cov}, this gives
\begin{equation}\label{eq:RS}
(\Delta F)^2(\Delta M)^2\ \ge\ (f,m)^2+\Big(\frac{\hbar}{2}\Big)^2\{f,m\}^2
=\Big(\tfrac12\ev{FM+MF}-\ev{F}\ev{M}\Big)^2+\frac14\big|\ev{[F,M]}\big|^2,
\end{equation}
which is the Robertson--Schr\"odinger uncertainty relation~\cite{Robertson,Schrodinger}. The two terms on the right-hand side are the metric and the symplectic products of the Hamiltonian vector fields, and the inequality expresses the fact that they are the real and imaginary parts of one Hermitian structure. Formula~(B5) of~\cite{v1} should be replaced by~\eqref{eq:RS}. See also~\cite{CMP,AS}.

\paragraph{The canonical commutation relations.} For a system whose classical phase space is $\R^{2N}$ with canonical coordinates $(q^i,p_j)$, the Hamiltonian vector fields of $q^i$ and $p_j$ generate translations and commute, $[X_{q^i},X_{p_j}]=-X_{\{q^i,p_j\}}=-X_{\delta^i_j}=0$, although $\{q^i,p_j\}=\delta^i_j\neq0$. In the quantum theory, $\hat q^i$ and $\hat p_j$ generate the Weyl unitary groups, and by~\eqref{eq:PB} the expectation-value functions satisfy exactly the same relations on $\Ps$:
\begin{equation}
\{\ev{\hat q^i},\ev{\hat p_j}\}=\delta^i_j,\qquad [X_{\ev{\hat q^i}},X_{\ev{\hat p_j}}]=0,
\end{equation}
the second because the commutator $i\hbar\,\delta^i_j\mathbb{1}$ generates only a phase, which projects to zero on $\Ps$ (on a suitable common invariant domain). Hence, at the level of the symplectic geometry of $\Ps$, the canonical pair behaves exactly as in classical mechanics, and the CCR are simply the statement $\{\ev{\hat q},\ev{\hat p}\}=1$. The uncertainty relation $\Delta q\,\Delta p\ge\hbar/2$ follows from~\eqref{eq:RS} with this Poisson bracket. What has no classical analogue is not the bracket but the metric. In classical mechanics states are points and have zero dispersion, whereas on $\Ps$ the dispersions are the $g$-lengths of the Hamiltonian vector fields, and Lemma~\ref{lem:kahler} converts the nonvanishing Poisson bracket into a lower bound on them. This sharpens the observation of~\cite{v1}: the uncertainty principle requires both geometric structures together with their compatibility.

\section{Discussion}\label{sec:discussion}

We have examined the superposition principle within the geometric formulation of quantum mechanics, in which the state space $\Ps$ is a nonlinear K\"ahler manifold. Our main proposal is that two physically distinct situations, which are usually described by the same words, must be distinguished.

In the first, exemplified by spin-$\frac12$ preparation, one has full control over the state, which may be polarized along any direction. This is the \emph{projective superposition principle}: any two states span a canonical, totally geodesic, area-minimizing sphere, whose points are their superpositions and on which all transition probabilities between them are computed. Its mathematical origin is the complex projective structure of $\Ps$, i.e.\ its family of lines, a remnant of the linearity of $\Hs$. We have shown that it has dynamical content: for $\dim\Hs\ge3$ the requirement that symplectic evolution preserve superpositions is equivalent to the linearity of the Schr\"odinger equation (Theorem~\ref{thm:dynamics}).

In the second, exemplified by the two-slit experiment, there is no freedom in the incoming state. The freedom lies in the wall, represented by a sum of orthogonal projectors. This is the \emph{decomposition principle}, and the linearity it relies on is that of the space of observables, i.e.\ of Killing vector fields on $\Ps$, not of the states. The interference pattern is invariant under the phase of the incoming state, because that state itself fixes the phase reference of the decomposition. We have shown that the pattern is completely geometric (Theorem~\ref{thm:interference}). The ``classical'' contributions of the slits are products of transition probabilities along geodesic paths, governed by $g$. The interference phase is a symplectic area, governed by $\omega$, and is well defined thanks to the integrality of $\omega/2\pi\hbar$. Thus the metric and the symplectic structure, which we found jointly responsible for the uncertainty relations, are also the two ingredients of interference.

These results suggest that the foundational content of the superposition principle can be phrased entirely in terms of the geometry of $\Ps$: the existence of the lines (preserved by the dynamics), the identification of transition probabilities with distances, and the identification of relative phases with symplectic areas. It would be interesting to know to what extent these properties single out complex projective space among candidate state spaces. Spaces in which the transition probability is a function of distance alone and the symmetries act transitively on pairs of points at equal distance are two-point homogeneous, and the compact ones form a very short list. Remarkably, this list coincides with the one obtained from operational postulates that include the absence of higher-order interference~\cite{BMU}. We hope to return to this question elsewhere.

\section*{Acknowledgments}
We thank J.A.~Zapata for discussions and M.~Gatti for correspondence. This work was partially supported by CONACyT grants J32754-E and U47857-F and DGAPA-UNAM grant 112401.

\appendix

\section{Geodesics and the lines of $\Ps$}\label{app:geodesic}

We prove Proposition~\ref{prop:sphere} and the totally geodesic property used in Proposition~\ref{prop:luders}. It suffices to work with $g_\FS$.

\paragraph{Geodesics.} The projection $\pi:\Sp\to\Ps$ is a Riemannian submersion with respect to the round metric $\mathrm{Re}\braket{\cdot}{\cdot}$ on $\Sp$ and $g_\FS$ on $\Ps$. Geodesics of $\Ps$ are therefore the projections of horizontal great circles~\cite{BZ,KN}: for $\psi\in\Sp$ and a unit vector $v$ with $\braket{\psi}{v}=0$,
\begin{equation}\label{eq:geod}
\gamma(t)=\big[\cos t\;\psi+\sin t\;v\big],
\end{equation}
is the unit-speed geodesic through $[\psi]$ with initial velocity $[v]$, and every geodesic is of this form. It has period $\pi$, and its length from $[\psi]$ to $\gamma(t)$, $0\le t\le\pi/2$, is $t=\arccos|\braket{\psi}{\cos t\,\psi+\sin t\,v}|$.

\paragraph{Proof of Proposition~\ref{prop:sphere}.} (a) A projective line is the projectivization of a two-dimensional subspace, and a two-dimensional subspace containing the independent vectors $\psi,\phi$ is their span.

(b) Every geodesic through $\Psi=[\psi]$ is of the form~\eqref{eq:geod} and is contained in the line $\Ps(\Span\{\psi,v\})$. If it also passes through $\Phi\neq\Psi$, then $\phi\in\Span\{\psi,v\}$, so this line is $L_{\Psi\Phi}$. Hence every geodesic of $\Ps$ joining two points of $L_{\Psi\Phi}$ lies in $L_{\Psi\Phi}$. This includes all minimizing geodesics, even when $\Psi\perp\Phi$ (then there is a circle of them, all in $L_{\Psi\Phi}$). Consequently the distance in $\Ps$ between two points of $L_{\Psi\Phi}$ equals their distance on $L_{\Psi\Phi}$. With the induced metric, $L_{\Psi\Phi}$ is the round sphere of radius $\tfrac12$: by~\eqref{eq:projsup}, the point with Bloch angles $(\beta,\alpha)$ is at distance $\beta/2$ from $\Psi$ and has $|\braket{\psi}{\cdot}|=\cos(\beta/2)$. This proves~\eqref{eq:prob}.

The same argument applies verbatim to $\Ps(W)$ for any subspace $W$: a geodesic with initial point and velocity in $\Ps(W)$ has $\psi,v\in W$ and stays in $\Ps(W)$. Alternatively, $\Ps(W)$ is a connected component of the fixed-point set of the isometry induced by the unitary $E_W-E_{W^\perp}$, and connected components of fixed-point sets of isometries are totally geodesic~\cite{Kobayashi}.

(c) The area of a round sphere of radius $\frac12$ is $\pi$. Since $\omega_\FS$ is a K\"ahler form, it is a calibration: for any oriented tangent two-plane $P$, $\omega_\FS|_P\le dA_P$, with equality iff $P$ is a complex line with its complex orientation (Wirtinger's inequality~\cite{HL}). Let $\Sigma$ be a closed oriented surface homologous to $L_{\Psi\Phi}$. Since $\omega_\FS$ is closed,
\[
\mathrm{Area}(\Sigma)\ \ge\ \int_\Sigma\omega_\FS=\int_{L_{\Psi\Phi}}\omega_\FS=\pi ,
\]
with equality iff $\Sigma$ is a complex curve. A compact complex curve in $\CP^{n-1}$ is algebraic (Chow's theorem~\cite{GH}), and its degree equals its homology class, so an irreducible curve of degree one is a projective line. Combined with (a), $L_{\Psi\Phi}$ is the unique minimizer containing $\Psi$ and $\Phi$. Other rational curves through two points, such as conics, have area $k\pi$ with $k\ge2$. \qed

\section{Bargmann invariants and symplectic area}\label{app:phase}

Let $\Psi_1,\dots,\Psi_m$ be states with $\braket{\psi_j}{\psi_{j+1}}\neq0$ (indices mod $m$), and let $\Gamma$ be the closed curve formed by the minimal geodesics $\Psi_1\to\Psi_2\to\dots\to\Psi_m\to\Psi_1$. We show that
\begin{equation}\label{eq:bargmann}
\arg\tr\big(\Pi_{\Psi_1}\Pi_{\Psi_2}\cdots\Pi_{\Psi_m}\big)=\arg\prod_{j}\braket{\psi_j}{\psi_{j+1}}=2\int_\Sigma\omega_\FS\pmod{2\pi}
\end{equation}
for any surface $\Sigma$ with $\partial\Sigma=\Gamma$~\cite{MS,SB,BZ}.

On $\Sp$ consider the one-form $A=\mathrm{Im}\braket{\psi}{d\psi}$, the Berry--Aharonov--Anandan connection~\cite{Berry,AharonovAnandan}. Its kernel on $T\Sp$ is the horizontal distribution, and \[dA(X,Y)=2\,\mathrm{Im}\braket{X}{Y}=2\,\omega_\FS(\pi_*X,\pi_*Y),\] since the vertical parts of $X,Y$ do not contribute to the imaginary part; thus $dA=\pi^*(2\,\omega_\FS)$.

\emph{Holonomy of geodesic polygons.} By~\eqref{eq:geod}, the horizontal lift of the minimal geodesic from $[\psi_j]$ to $[\psi_{j+1}]$ starting at $\psi_j$ ends at $e^{-i\alpha_j}\psi_{j+1}$, $\alpha_j=\arg\braket{\psi_j}{\psi_{j+1}}$, the representative \emph{in phase} with $\psi_j$ in Pancharatnam's sense~\cite{Pancharatnam}. Since horizontal lifts commute with phase rotations, the horizontal lift $\tilde\Gamma$ of $\Gamma$ starting at $\psi_1$ ends at $e^{-i\gamma}\psi_1$, with $\gamma=\sum_j\alpha_j=\arg\prod_j\braket{\psi_j}{\psi_{j+1}}$.

\emph{Holonomy as an area.} We may take $\Sigma$ to be a disk, over which the bundle $\Sp\to\Ps$ has a section $s$. Write $\tilde\Gamma(t)=e^{i\varphi(t)}s(\Gamma(t))$. Since $A$ changes under $\psi\mapsto e^{i\varphi}\psi$ by $d\varphi$, horizontality gives $\dot\varphi=-(s^*A)(\dot\Gamma)$. Therefore, by Stokes' theorem,
\[
-\gamma\equiv\varphi(1)-\varphi(0)=-\oint_\Gamma s^*A=-\int_\Sigma s^*dA=-2\int_\Sigma\omega_\FS \pmod{2\pi},
\]
which is~\eqref{eq:bargmann}. A general surface with boundary $\Gamma$ differs from a disk by a closed surface, which changes $2\int\omega_\FS$ by a multiple of $2\pi$. As a check on signs and orientations, we have verified~\eqref{eq:bargmann} numerically for random triangles in $\CP^2$ and quadrilaterals in $\CP^5$, by direct integration of $\omega_\FS$ over geodesic cones. Theorem~\ref{thm:interference} is the case $m=4$, with $(\Psi_1,\Psi_2,\Psi_3,\Psi_4)=(X,A_1,\Psi,A_2)$.


\begin{thebibliography}{99}

\bibitem{v1} A.~Corichi, ``Quantum superposition principle and geometry,'' arXiv:quant-ph/0407242v3 (2006); a version appeared in Gen.\ Rel.\ Grav.\ \textbf{38}, 677--687 (2006).

\bibitem{Kibble} T.~W.~B.~Kibble, ``Geometrization of quantum mechanics,'' Commun.\ Math.\ Phys.\ \textbf{65}, 189 (1979).

\bibitem{Heslot} A.~Heslot, ``Quantum mechanics as a classical theory,'' Phys.\ Rev.\ D \textbf{31}, 1341 (1985).

\bibitem{CMP} R.~Cirelli, A.~Mani\`a and L.~Pizzocchero, ``Quantum mechanics as an infinite-dimensional Hamiltonian system with uncertainty structure,'' J.\ Math.\ Phys.\ \textbf{31}, 2891 (1990).

\bibitem{AS} A.~Ashtekar and T.~A.~Schilling, ``Geometrical formulation of quantum mechanics,'' in \emph{On Einstein's Path}, ed.\ A.~Harvey (Springer, New York, 1998) [arXiv:gr-qc/9706069].

\bibitem{BH} D.~C.~Brody and L.~P.~Hughston, ``Geometric quantum mechanics,'' J.\ Geom.\ Phys.\ \textbf{38}, 19 (2001) [arXiv:quant-ph/9906086].

\bibitem{BZ} I.~Bengtsson and K.~\.Zyczkowski, \emph{Geometry of Quantum States}, 2nd ed.\ (Cambridge University Press, Cambridge, 2017).

\bibitem{AA90} J.~Anandan and Y.~Aharonov, ``Geometry of quantum evolution,'' Phys.\ Rev.\ Lett.\ \textbf{65}, 1697 (1990).

\bibitem{Dirac} P.~A.~M.~Dirac, \emph{The Principles of Quantum Mechanics}, 4th ed.\ (Clarendon Press, Oxford, 1958).

\bibitem{Feynman} R.~P.~Feynman, R.~B.~Leighton and M.~Sands, \emph{The Feynman Lectures on Physics}, Vol.~III (Addison-Wesley, Reading, 1965).

\bibitem{Luders} G.~L\"uders, ``\"Uber die Zustands\"anderung durch den Messproze\ss,'' Ann.\ Phys.\ (Leipzig) \textbf{8}, 322 (1951).

\bibitem{MMSZ} V.~I.~Man'ko, G.~Marmo, E.~C.~G.~Sudarshan and F.~Zaccaria, ``Inner composition law of pure states as a purification of impure states,'' Phys.\ Lett.\ A \textbf{273}, 31 (2000).

\bibitem{CGM1} R.~Cirelli, M.~Gatti and A.~Mani\`a, ``On the nonlinear extension of quantum superposition and uncertainty principles,'' J.\ Geom.\ Phys.\ \textbf{29}, 64 (1999).

\bibitem{CGM2} R.~Cirelli, M.~Gatti and A.~Mani\`a, ``The pure state space of quantum mechanics as Hermitian symmetric space,'' J.\ Geom.\ Phys.\ \textbf{45}, 267 (2003).

\bibitem{Weinberg} S.~Weinberg, ``Testing quantum mechanics,'' Ann.\ Phys.\ (N.Y.) \textbf{194}, 336 (1989).

\bibitem{Artin} E.~Artin, \emph{Geometric Algebra} (Interscience, New York, 1957).

\bibitem{Wigner} E.~P.~Wigner, \emph{Group Theory and its Application to the Quantum Mechanics of Atomic Spectra} (Academic Press, New York, 1959).

\bibitem{Bargmann} V.~Bargmann, ``Note on Wigner's theorem on symmetry operations,'' J.\ Math.\ Phys.\ \textbf{5}, 862 (1964).

\bibitem{Gleason} A.~M.~Gleason, ``Measures on the closed subspaces of a Hilbert space,'' J.\ Math.\ Mech.\ \textbf{6}, 885 (1957).

\bibitem{Uhlhorn} U.~Uhlhorn, ``Representation of symmetry transformations in quantum mechanics,'' Ark.\ Fys.\ \textbf{23}, 307 (1963).

\bibitem{Gisin} N.~Gisin, ``Weinberg's non-linear quantum mechanics and supraluminal communications,'' Phys.\ Lett.\ A \textbf{143}, 1 (1990).

\bibitem{Polchinski} J.~Polchinski, ``Weinberg's nonlinear quantum mechanics and the Einstein-Podolsky-Rosen paradox,'' Phys.\ Rev.\ Lett.\ \textbf{66}, 397 (1991).

\bibitem{MS} N.~Mukunda and R.~Simon, ``Quantum kinematic approach to the geometric phase. I. General formalism,'' Ann.\ Phys.\ (N.Y.) \textbf{228}, 205 (1993).

\bibitem{SB} J.~Samuel and R.~Bhandari, ``General setting for Berry's phase,'' Phys.\ Rev.\ Lett.\ \textbf{60}, 2339 (1988).

\bibitem{Sorkin} R.~D.~Sorkin, ``Quantum mechanics as quantum measure theory,'' Mod.\ Phys.\ Lett.\ A \textbf{9}, 3119 (1994).

\bibitem{Sinha} U.~Sinha, C.~Couteau, T.~Jennewein, R.~Laflamme and G.~Weihs, ``Ruling out multi-order interference in quantum mechanics,'' Science \textbf{329}, 418 (2010).

\bibitem{BMU} H.~Barnum, M.~P.~M\"uller and C.~Ududec, ``Higher-order interference and single-system postulates characterizing quantum theory,'' New J.\ Phys.\ \textbf{16}, 123029 (2014).

\bibitem{Robertson} H.~P.~Robertson, ``The uncertainty principle,'' Phys.\ Rev.\ \textbf{34}, 163 (1929).

\bibitem{Schrodinger} E.~Schr\"odinger, ``Zum Heisenbergschen Unsch\"arfeprinzip,'' Sitzungsber.\ Preuss.\ Akad.\ Wiss., Phys.-Math.\ Kl.\ \textbf{14}, 296 (1930).

\bibitem{KN} S.~Kobayashi and K.~Nomizu, \emph{Foundations of Differential Geometry}, Vol.~II (Interscience, New York, 1969).

\bibitem{Kobayashi} S.~Kobayashi, ``Fixed points of isometries,'' Nagoya Math.\ J.\ \textbf{13}, 63 (1958).

\bibitem{HL} R.~Harvey and H.~B.~Lawson, ``Calibrated geometries,'' Acta Math.\ \textbf{148}, 47 (1982).

\bibitem{GH} P.~Griffiths and J.~Harris, \emph{Principles of Algebraic Geometry} (Wiley, New York, 1978).

\bibitem{Sontz1} S.~B.~Sontz, ``Quantum probability geometrically realized in projective space,'' Found.\ Phys.\ \textbf{55}, 62 (2025).

\bibitem{Sontz2} S.~B.~Sontz, ``All quantum probability viewed in complex projective geometry,'' arXiv:2605.17578 (2026).

\bibitem{Sontz3} S.~B.~Sontz, ``A new definition of quantum superposition,'' arXiv:2606.15607 (2026), submitted to SciPost Phys.\ Proc.

\bibitem{Pancharatnam} S.~Pancharatnam, ``Generalized theory of interference, and its applications,'' Proc.\ Indian Acad.\ Sci.\ A \textbf{44}, 247 (1956).

\bibitem{Berry} M.~V.~Berry, ``Quantal phase factors accompanying adiabatic changes,'' Proc.\ R.\ Soc.\ Lond.\ A \textbf{392}, 45 (1984).

\bibitem{AharonovAnandan} Y.~Aharonov and J.~Anandan, ``Phase change during a cyclic quantum evolution,'' Phys.\ Rev.\ Lett.\ \textbf{58}, 1593 (1987).

\end{thebibliography}
\end{document}